\documentstyle[aps,prb,epsf]{revtex}

\newcommand{\be}{\begin{equation}}
\newcommand{\ee}{\end{equation}}
\newcommand{\bea}{\begin{eqnarray}}
\newcommand{\eea}{\end{eqnarray}}
\newcommand{\hh}{\hbar_{\rm eff}}
\newcommand{\cH}{{\cal H}}

\begin{document}
\twocolumn[\hsize\textwidth\columnwidth\hsize\csname @twocolumnfalse\endcsname
\title{Metal-insulator transitions in cyclotron resonance 
of periodic nanostructures due to avoided band crossings}

\author{L.~Hufnagel$^1$, M.~Weiss$^1$, A.~Iomin$^2$, R.~Ketzmerick$^1$, 
S.~Fishman$^2$, and T.~Geisel$^1$ \\
${}^1$ Max-Planck-Institut f\"ur Str\"omungsforschung and 
Institut f\"ur Nichtlineare Dynamik der Universit\"at G\"ottingen, \\
Bunsenstr. 10, 37073 G\"ottingen, Germany \\
${}^2$ Department of Physics, Technion, Haifa 32000, Israel}
\maketitle

\begin{abstract}
A recently found metal--insulator transition in a model for cyclotron resonance
in a two-dimensional periodic potential is investigated by means of spectral 
properties of the time evolution operator.
The previously found dynamical signatures of the transition are explained in terms 
of avoided band crossings due to the change of the external electric field.
The occurrence of a cross-like transport is predicted and numerically confirmed.
\end{abstract}

]

%----------------------------------------------------------------------------------------
%
\section{\bf Introduction}
Since the pioneering work of Bloch~\cite{bloch29} and Landau~\cite{landau30} the 
investigation of spectral properties and transport features of Bloch electrons in 
magnetic fields~\cite{peierls} have attracted much attention. While infinitely degenerate
Landau levels are found for free electrons in a static magnetic field, a spatially periodic 
potential without field yields a band structure in the spectrum. The general situation 
of a periodic potential {\it and} a magnetic field is still a matter of intensive 
research\cite{review1,review2,review3,KKSG00}. A particular simple realization 
for a two-dimensional (2D) periodic potential is, e.g., given by
\be\label{potential}
V({\bf r})=V_x\cos\left(\frac{2\pi x}{a}\right)+
V_y\cos\left(\frac{2\pi y}{b}\right),
\ee
where $a,\,b $ are the periods in the corresponding directions.
In a one-band approximation this potential leads to the well-known Harper 
model~\cite{harper55}, where a metal--insulator transition (MIT) is 
observed upon changing the modulation amplitudes from $V_y>V_x$ to 
$V_x>V_y $~\cite{AA79}. At the critical point, i.e., $V_x=V_y$, the spectrum and 
the eigenfunctions are multifractals~\cite{hof76} and give rise to 
anomalous diffusion~\cite{HA88,KKKG97,GKP91b}. The one-band Harper model, however, has an 
integrable classical limit, whereas Bloch electrons in a magnetic field show chaotic 
dynamics in the classical limit. In quantum mechanics this is reflected in the coupling 
of Landau bands. This coupling leads to avoided band crossings, which induce an unusual transport
phenomenon: For~$V_x\neq V_y$ electrons show ballistic transport in the direction of strong
potential modulation and localization in the direction of weaker modulation~\cite{KKSG00}.

Analogous phenomena can be found in driven systems with a chaotic classical limit even within
the one-band approximation. A well studied example is the kicked Harper model 
(KHM)~\cite{LS91,GKP91a,ACS92,dana94,KKG98,iomin}.
For small kicking strength its behavior is analogous to the Harper model~\cite{GKP91a}, while for 
increasing
kicking strength avoided band crossings due to the classical non-integrability induce a variety
of MITs~\cite{KKG98}. Recently, it was demonstrated that a variety of kicked models, including the 
KHM, can be realized for electrons on a lattice in the presence of a magnetic field driven by
a smooth electric field~\cite{IF99}. The effective kicking is a result of a resonant 
interaction between the electronic motion and the driving field. These models correspond
to cyclotron resonance experiments in antidot arrays and in organic metals~\cite{IF99,IF00}.
By investigating the dynamics, a localization-delocalization 
transition induced by the amplitude of the electric field was observed~\cite{IF00} 
and remained unexplained so far.

In this paper we investigate these dynamical properties in terms of spectral properties
in a wide range of driving strengths. We show that the previously found localization-delocalization
transition is a consequence of avoided band crossings due to the change of the driving field strength.
Moreover, the spectral analysis predicts a cross-like ballistic transport, which is numerically
confirmed. We start in Section~II by briefly introducing the model and explain
the numerical approach in Section~III. In Section~IV we present our main results, 
which give a clear picture for the MIT in terms of the spectrum. We finally summarize 
our findings in Section~V.

%----------------------------------------------------------------------------------------
%
\section{The model}
The one particle Hamiltonian, describing the cyclotron resonance of an electron
in the 2D periodic potential~$V({\bf r})$ of Eq.~(\ref{potential}) can be written 
in the following form:
\begin{equation}\label{H1}
\cH=\frac{1}{2m^*}\left({\bf p}-e{\bf A}({\bf r},t)\right)^2+
V({\bf r})\,\,,
\end{equation}
where $ {\bf p}=(p_x,p_y) $ is the momentum of an electron with effective mass~$m^*$ 
and charge~$e$ moving in the $xy$-plane perpendicular to the constant magnetic field $B$
pointing in $z$-direction. We choose the Landau gauge, which yields a vector 
potential ${\bf A}({\bf r},t)=(\frac{E_x}{\nu}\sin\nu t, xB-\frac{E_y}{\nu}\cos\nu t)$ 
containing also the alternating electric field ${\bf E}=(-E_x\cos\nu t,\,E_y\sin\nu t) $ 
with frequency~$\nu$. 

Applying a gauge transformation to the Hamiltonian~(\ref{H1}), the time dependence can be 
shifted to the periodic potential~$V({\bf r})\rightarrow V({\bf r},t)$~\cite{IF00}: If the 
fields~$B,~E_x,~E_y$ satisfy $\omega^*E_y=\nu E_x $ then one can rewrite the 
Hamiltonian as~$\cH=\cH_0({\bf p},x)+V({\bf r},t)$, where $\cH_0=\cH_0({\bf p},x) $ describes the 
cyclotron motion with frequency~$\omega^*=e B/m^*$. The perturbation~$V({\bf r},t)$ splits the
infinitely degenerate Landau levels of~$\cH_0$ into a mini-band structure. For strong electric 
fields and~$\nu=1$ this Hamiltonian~$\cH $ reduces to a kicked system~\cite{IF99,IF00}
\bea\label{hamiltonKKHM}
\tilde{\cH}=L\cos p+ K\sum_{n=-\infty}^{\infty}&&\left\{ \right.
\cos(q-\kappa_0)\delta(t-2n+1)\nonumber\\
&+& \cos(q+\kappa_0)\delta(t-2n)  \left.\right\}\,\, ,
\eea
where the canonical operators~$q$ and~$p$ corresponding to the spatial coordinates $x,y$ 
satisfy $[p,q]=-i2\pi\hh$ and $\kappa_0=(2\pi eE_y/(bm^*)-\pi/4)~{\rm mod}~{2\pi}$;
$L,\,K$ denote the rescaled amplitudes of the potential~(\ref{potential}).
The effective Planck constant~$\hh=\hbar/(abeB)$,
corresponds to the inverse number of magnetic flux quanta through a unit 
cell~$a\times b$. The system described by Eq.~(\ref{hamiltonKKHM}) has a chaotic 
classical limit~\cite{IF99} and for~$\kappa_0=0$ reduces to the KHM. 

%----------------------------------------------------------------------------------------
%
\section{\bf Numerical Approach}
The time evolution operator for one period corresponding to~(\ref{hamiltonKKHM}) reads
\bea\label{KKHM}
{\cal U}&=&\exp\left(-\frac{iL}{\hh}\cos p\right)
\exp\left(-\frac{i K}{\hh}\cos(q-\kappa_0)\right)\nonumber\\
&&\times \exp\left(-\frac{iL}{\hh}\cos p\right)\exp\left(-\frac{i K}{\hh}\cos(q+\kappa_0)\right)\,\, .
\eea
In order to obtain its spectrum, one has to approximate the irrational effective 
Planck constant by a rational approximant, $\hh=2\pi M/N$. 
Following standard techniques for kicked systems~\cite{CS86}, this leads to a 
$N\times N$-matrix for the evolution operator~$\cal U$, which depends on two Bloch 
phases~$\theta_q,\theta_p\in[0,2\pi]$. These Bloch phases arise from periodic boundary conditions
in the~$q$- and $p$-direction, respectively. 
Diagonalization of~$\cal U$ for all pairs~$(\theta_q,\theta_p)$ gives its spectrum.
Keeping~$\theta_q$ fixed and varying~$\theta_p$ yields a part of this spectrum, from now
on referred to as the ``$p$-spectrum''.
Eigenfunctions, which are localized in $p$-direction will be associated
with very narrow bands in the $p$-spectrum, whereas extended states give rise to broad bands. 
Since for higher approximants of~$\hh$ the width of the narrow bands decreases 
exponentially, we will call them levels.
In an analogous way one can define the ``$q$-spectrum'', which reflects the properties of the
eigenfunctions in $q$-direction. Figure~\ref{fig:spek1} shows a $q$- and $p$-spectrum, where
one observes broad bands and levels.

In order to investigate dynamics, we will study the spreading of wave packets, initially 
$\delta$-localized in either $q$- or $p$-direction by determining the corresponding variances
$M_q(t)$ and $M_p(t)$. For the spreading in $p$-direction we use periodic boundary conditions 
in $q$-direction
after one unit cell, giving a discrete lattice in momentum space with spacing~$\hh$.
An analogous construction is used for the spreading in $q$-direction. When considering transport
in phase space in both directions, we use periodic boundary conditions after $r$ unit cells in 
$q$-direction, which gives $r$ different Bloch phases $\theta_q=2\pi j/r$ ($j=0,\ldots,r-1$).
The complete wave packet is then obtained by combining the wave packets for each Bloch phase and 
it is defined on a grid with spacing~$\hh/r$ in momentum space.

%----------------------------------------------------------------------------------------
%
\section{\bf Results}
We will study the spectrum of Eq.~(\ref{KKHM}) as a function of~$\kappa_0$ and its 
consequences for dynamical properties. In particular, we will focus on the parameters
used in Ref.~\cite{IF00} and show, that they reflect the generic behavior of the 
model~(\ref{hamiltonKKHM}). 

In Ref.~\cite{IF00} for $K=L=5 and \hh=2\pi/(7+\sigma_g)$, 
where $\sigma_g=(\sqrt{5}-1)/2$ denotes the golden mean, a transition from diffusive to
ballistic dynamics was found, when~$\kappa_0$ was varied from~$0$ to~$\pi\sigma_g$
(cf. Fig.~1 of Ref.~\cite{IF00}). In Fig.~\ref{fig:spek1} we present the $q$- and $p$-spectrum for 
one period of $\kappa_0$. Apparently, the duality of $q$- and $p$-spectrum is conserved for all~$\kappa_0$, 
i.e., the $q$-spectrum shows wide bands, where the $p$-spectrum is level-like and vice versa. 

For~$\kappa_0=0$ the variances increase almost linearly in both $q$- and $p$-direction, 
corresponding to diffusive spreading in either direction (top of Fig.~\ref{fig:dyn1}a). This well-known
behavior of the KHM is a direct consequence of the multifractal nature of spectrum and 
eigenfunctions\cite{LS91,GKP91a,ACS92}.

For~$\kappa_0=\pi\sigma_g$ bands occur in the $p$-spectrum explaining the ballistic transport in 
$p$-direction observed in Ref.~\cite{IF00}. There are, however, also levels, which are related to 
bands in the $q$-spectrum, predicting a ballistic spreading in $q$-direction. This is confirmed 
in Fig.~\ref{fig:dyn1}b. All eigenstates are extended in either of the directions and 
localized in the other one.
A typical initial wave packet will excite both types of eigenfunctions.
The part of the wave packet consiting of eigenfunctions extended in the $q$-direction
gives ballistic transport in the $q$-direction. The other part, which is a superposition
of eigenfunctions extended in the $p$-direction, yields ballistic transport in the $p$-direction.
Therefore one finds a superposition of ballistic transport in $q$- and $p$-direction,
namely a cross-like transport (bottom of Fig.~\ref{fig:dyn1}b).
The cross-like form of the wave packet is clearly different from the 
isotropic form for~$\kappa_0=0$ (Fig.~\ref{fig:dyn1}a).
There are several other transport features one can observe from the Husimi plot in Fig.~\ref{fig:dyn1}b.
The weight of the wave packet spreading ballistically along the $p$-direction
is much bigger than the corresponding weight for the $q$-direction.
This is a signature of the spectrum, namely that there are more bands in the $p$-spectrum than
there are bands in the $q$-spectrum at $\kappa_0=\pi\sigma_g$, as can be seen in Fig.~\ref{fig:spek1}.
It should be noted, that the velocities in $p$- and $q$-direction (which can be inferred from the
variances $M_p(t)$ and $M_q(t)$) are in general not related to the number of bands but are determined
by the band widths.

Now we will turn away from the symmetry line $K=L$ and focus on the parameters
$K=2,\,L=4,\hh=2\pi/(7+\sigma_g)$. In Fig.~\ref{fig:spek2} we show the
$q$- and $p$-spectrum as a function
of~$\kappa_0$, which again appears to be dual.
For the KHM ($\kappa_0=0$) the $q$-spectrum shows broad bands, whereas the $p$-spectrum is level-like, 
indicating transport in $q$-direction, but localization in $p$-direction\cite{LS91}. This 
is confirmed by the dynamics (Fig.~\ref{fig:dyn2}a). Variation of~$\kappa_0$ induces avoided band 
crossings leading to 
the occurrence of bands in the $p$-spectrum (see magnification in Fig.~3). This is associated 
with a localization-delocalization transition in $p$-direction as observed in 
Ref.~\cite{IF00}.  As there are still bands in the $q$-spectrum one again finds 
the generic cross-like transport (Fig.~\ref{fig:dyn2}b).
Here, the weight of the wave packet spreading along the $q$-direction is bigger than the
corresponding weight in $p$-direction, as is implied by the much larger number of bands in
the $q$-spectrum (Fig.~\ref{fig:spek2}).
Another characteristic of the wave packet is the increased localization length in $p$-direction
of the part of the wave packet spreading in $q$-direction compared to Fig.~\ref{fig:dyn2}a.
This is due to the increased localization length of the eigenstates corresponding to the levels
in the $p$-spectrum.
This change in the localization length can be explained by avoided band crossings\cite{KKG98}.

%----------------------------------------------------------------------------------------
%
\section{\bf Summary}
In conclusion, we have shown here, that the previously found MIT in cyclotron 
resonance in a 2D-periodic potential in the presence of a magnetic field can be fully understood by 
investigating the spectrum of the corresponding time evolution operator. 
We have presented numerical evidence, that the duality of the spectra is conserved under
changes of the driving strength.
We were able to explain the previously observed dynamics~\cite{IF00}, 
which has shown a transition from diffusive to ballistic
transport as well as a transition from localization to ballistic transport in 
the $p$-direction. Furthermore, the spectra allowed us to predict the transport
behavior in $q$-direction leading to a cross-like transport. These predictions
were also confirmed numerically.
We would like to mention, that cross-like transport is generic for systems with dual spectra
and a chaotic classical limit.

The observed transport phenomena in phase space of the effective Hamiltonian~(\ref{hamiltonKKHM})
correspond to analogous transport phenomena in the $xy$-plane.
The required conditions for realization of the observed MIT may well be achieved 
in experiments on cyclotron resonance~\cite{vas} of a 2D electron gas embedded 
in lateral superlattices fabricated on GaAs heterostructures~\cite{ccoo11}. 

This research was supported by the Niedersachsen Ministry of Science. A.I. and S.F.
were also supported by the Israel Science Foundation founded by the Israel Academy of 
Sciences and Humanities. A.I. thanks Prof.~T.~Geisel for his hospitality at the MPI f\"ur
Str\"omungsforschung.
%----------------------------------------------------------------------------------------
%

\begin{figure}
\begin{center}
\epsfxsize=7.6cm
\leavevmode
%    \epsffile{fig1.eps}
\caption{
The $p$-spectrum (top) and $q$-spectrum (bottom) of the evolution
operator (\ref{KKHM}) for $K=L=5$ vs.~$\kappa_0$ (one period $[-\pi/2,\pi/2]$ is shown). 
The approximant for the irrational effective Planck constant is~$\hh=2\pi\cdot 34/259$.
Arrows denote the values $\kappa_0=0$ and $\kappa_0=\pi\sigma_g$  
used in Fig.~\ref{fig:dyn1}. For~$\kappa_0=0$ Eq.~(\ref{hamiltonKKHM}) reduces to the KHM and
the spectrum is multifractal, while for~$\kappa_0\neq 0$ it consists of bands and levels.
The spectra appear to be dual to each other,  since broad bands in one of them are associated 
with levels in the other.
}\label{fig:spek1}
\end{center}
\end{figure}

\begin{figure}
\begin{center}
\epsfxsize=7.6cm
\leavevmode
%    \epsffile{fig2.eps}
\caption{(a) Variances $M_q(t)$, $M_p(t)$ and the Husimi plot of a 2D wave packet after 100 kicks
(one point per unit cell and $r=89$ cells in each direction)
for $K=L=5$, $\hh=2\pi/(7+\sigma_g)$, $\kappa_0=0$.
(b) same for~$\kappa_0=\pi\sigma_g$.
In the first case, the variances show a diffusive-like behavior while
the wave packet spreads ballistically in both directions in the second case.
This is also reflected in the Husimi plots, where one observes that the wave packet spreads 
isotropically in phase space in (a) and shows a cross-like spreading in (b).
}\label{fig:dyn1}
\end{center}
\end{figure}

\begin{figure}
\begin{center}
\epsfxsize=7.6cm
\leavevmode
%    \epsffile{fig3.eps}
\caption{
Same as in Fig.~\ref{fig:spek1} for $K=2,\,L=4$.
For $\kappa_0=0$ the $p$-spectrum consists of levels only and the $q$-spectrum shows bands.
Magnifications ($\hh = 2\pi 55/419$) show part of the spectra around~$\kappa_0=\pi\sigma_g$,
where one finds bands in the $p$-spectrum and levels in the $q$-spectrum.
Duality again seems to be conserved.
}\label{fig:spek2}
\end{center}
\end{figure}

\begin{figure}
\begin{center}
\epsfxsize=7.6cm
\leavevmode
%    \epsffile{fig4.eps}
\caption{
Same as in Fig.~\ref{fig:dyn1} for $K=2,\, L=4$.
(a) Ballistic spreading in  $q$-direction and localization in $p$-direction is observed.
(b) Ballistic spreading in both directions leads to a cross-like transport.
}\label{fig:dyn2}
\end{center}
\end{figure}


\begin{thebibliography}{99}

\bibitem{bloch29} F.~Bloch, Z.~Phys. {\bf 52}, 555 (1929).
%
\bibitem{landau30} L.~Landau, Z.~Phys. {\bf 64}, 629 (1930).
%
\bibitem{peierls} R.~Peierls, Z.~Phys. {\bf 80}, 763 (1933).
%
\bibitem{review1} C.~Albrecht {\em et al.}, Phys. Rev. Lett. {\bf 93}, 2234 (1999).
%
\bibitem{review2} T.~Schl\"osser {\em et al.}, Europhys. Lett. {\bf 33}, 683 (1996).
%
\bibitem{review3} A.~Barelli, J.~Bellisard, and F.~Claro, Phys. Rev. Lett. {\bf 83}, 5082 (1999).
%
\bibitem{KKSG00} R.~Ketzmerick, K.~Kruse, D.~Springsguth, and T.~Geisel, Phys. Rev. 
Lett. {\bf 84}, 2929 (2000).
%
\bibitem{harper55} P.G.~Harper, Proc.~Phys.~Soc.~Lond. {\bf A68}, 874 (1955).
%
\bibitem{AA79} S.~Aubry and G.~Andr\'e, in Proc. of the Israel Phys.
Soc., edited by C. G. Kuper, v. 3, 133 (Hilger, Bristol, 1979).
%
\bibitem{hof76} D.~R.~Hofstadter, Phys. Rev. {\bf B 14}, 2239 (1976);
C.~Tang and M.~Kohmoto, Phys. Rev. {\bf B 34}, 2041 (1986).
%
\bibitem{HA88} H.~Hiramoto and S.~Abe, J.~Phys.~Soc.~Jpn. {\bf 57}, 230
and 1365 (1988).
%
\bibitem{GKP91b} T.~Geisel, R.~Ketzmerick, and G.~Petschel, Phys.~Rev.~Lett. {\bf 66}, 1651 (1991).
%
\bibitem{KKKG97} R.~Ketzmerick, K.~Kruse, S.~Kraut, and T.~Geisel,
Phys.~Rev.~Lett. {\bf 79}, 1959 (1997).
%
\bibitem{LS91} R.~Lima and D.~Shepelyansky, Phys.~Rev.~Lett. {\bf 67}, 1377 (1991).
%
\bibitem{GKP91a} T.~Geisel, R.~Ketzmerick, and G.~Petschel, Phys.~Rev.~Lett. {\bf 67}, 3635 (1991);
{\bf 69}, 695 (1992).
%
\bibitem{ACS92} R.~Artuso, G.~Casati, and D.~Shepelyansky,  Phys. Rev. Lett. {\bf 68}, 3826 (1992); 
R.~Artuso {\it et al.},  Phys.~Rev.~Lett. {\bf 69}, 3302 (1992).
%
\bibitem{dana94} I.~Dana, Phys.~Rev.~Lett. {\bf 73}, 1609 (1994); Phys.~Lett. {\bf A 197}, 413 (1995).
%
\bibitem{KKG98} R.~Ketzmerick, K.~Kruse, and T.~Geisel, Phys.~Rev.~Lett. {\bf 80}, 137 (1998).
%
\bibitem{iomin} A.~Iomin, G.M.~Zaslavsky, Phys.~Rev.~E {\bf 60}, 7580 (1999); Chaos {\bf 10}, 147 (2000).
%
\bibitem{IF99} A.~Iomin and S.~Fishman, Phys.~Rev.~Lett. {\bf 81},
1921 (1998); Physica {\bf D 131}, 170 (1999).
%
\bibitem{IF00}  A.~Iomin and S.~Fishman, Phys.~Rev.~B {\bf 61}, 2085 (2000).
% 
\bibitem{CS86} S.-J.~Chang, K.-J.~Shi, Phys.~Rev.~A {\bf 34}, 7 (1986).
%
\bibitem{vas} E.~Vasiliadou, {\em et al}., Phys. Rev. {\bf B 52}, R8658 (1995).
%
\bibitem{ccoo11} For a review, see W.~Hansen, U.~Merkt, and J.P.~Kotthaus, 
in {\em Nanostructured Systems}, edited by M.~Reed, Semiconductors and Semimetals Vol.~35 
(Academic, San Diego, 1992), p. 279.

\end{thebibliography}
\end{document}